\documentstyle[prl,aps,twocolumn,psfig]{revtex}

\begin{document}

\twocolumn[\hsize\textwidth\columnwidth\hsize\csname
@twocolumnfalse\endcsname

\draft

\title{Forbidden activation levels in a non-stationary tunneling process}
\author{Er'el Granot \footnotemark}

\address{School of Physics and Astronomy, Raymond and Beverly Sackler Faculty of Exact Sciences, Tel-Aviv University, 69978 Tel Aviv, Israel}

\date{\today}
\maketitle
\begin{abstract}
\begin{quote}
\parbox{16 cm}{\small
Tunneling in the presence of an opaque barrier, part of which
varies in time, is investigated numerically and analytically in
one dimension. Clearly, due to the varying barrier a tunneling
particle experiences spectral widening. However, in the case of
strong perturbations, the particles' activation to certain
energies is avoided. We show that this effect occurs only when the
perturbation decays faster than $t^{-2}$ }
\end{quote}
\end{abstract}


]

\narrowtext \footnotetext{erel.g@kailight.com} \noindent

Tunneling, which is a very common phenomenon in the quantum world,
is rarely a stationary process. In most cases, it is involved with
temporal changes of the medium through which the particle is
tunneling. In some cases, such variations are mostly due to
transport of the tunneling particles themselves. In these cases,
the problems are intrinsically nonlinear. However, when the source
of the changes is external, the problem can be treated as a linear
one. In this category, one can even find stationary many-body
transport problems where a single particle can experience the
influence of its surroundings (in this case its neighbors) as an
alternating external force. Many works (see, for example, refs.
\cite{Zangwill,Ivlev_Melnikov,Fisher,Azbel,Aleiner_Andreev}) done
in this field have revealed an abundance of physical effects:
resonances and oscillations in energy, exponential increase of the
tunneling current, activation-assisted tunneling and elevator
resonance activation.

This paper discusses the influence of a very strong perturbation
(compared with the incoming particles' energy) on the tunneling
process. It is assumed that the perturbation potential is always
positive (see Fig.1), i.e., it increases the barrier (locally) and
never creates a local well (inside the barrier). Therefore,
effects like elevator resonance activation do not occur (see ref.
\cite{Azbel}). On the contrary, in the adiabatic approximation,
the perturbation - being very strong - should reduce the
transmission dramatically. However, when the characteristic time
of the perturbation decreases, the particle may absorb energy from
the external perturbation, which can assist it in its due course
of tunneling. It is clear, for example, that in the extreme case,
where the potential barrier is smaller than $\hbar/\tau$ ($\tau$
is the characteristic time of the perturbation), the particle can
easily (with high probability) absorb enough energy to leap over
the barrier. This paper discusses the intermediate case, where the
process has a tunneling nature but allows for energy quanta to be
absorbed by the tunneling particle. A very strong perturbation,
which is activated for a finite period, permits propagation only
at the very onset and at the final decay of the perturbation. When
the wave function in these two instances has the same phase,
maximum transmission is expected, but when in these two moments
the wave function is out of phase, one should expect minimum
transmission. These are the forbidden activation energies we are
looking for.

\begin{figure}
\psfig{figure=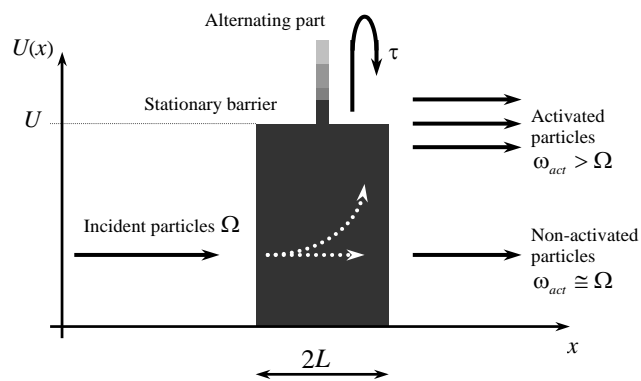,width=10cm,bbllx=140bp,bblly=160bp,bburx=700bp,bbury=500bp,clip=}
\caption{Activation in a tunneling process due to a varying
potential barrier. The different gray levels represent the
temporally varying part of the potential.}\label{fig1}
\end{figure}

This system, which is presented in Fig.1, can be described in
terms of the Schr\"{o}dinger equation in the following way

\begin{equation}
-\frac{d^2 \psi}{dx^2}+U(x)\psi+f(t/\tau)\delta(x) \psi=i
\frac{d\psi}{dt} \label{Schro_eq}
\end{equation}

where we have used the units $\hbar=1$ and $2m=1$ ($m$ is the
electron's mass), and $f(x)$ is a function, which vanishes very
rapidly for $x\rightarrow \infty$, and $U(x)$ is the barrier's
potential.

We seek the solution
\begin{equation}
\psi(x,t)=\psi_{\Omega}(x,t)+\int d\omega e^{-i\omega
t}a_{\omega}G_{\omega}^+(x) \label{seek_solution}
\end{equation}

where $\psi_{\Omega}(x,t)$ is the homogeneous solution of
eq.\ref{Schro_eq}, $\psi_{\Omega}(x,t)\equiv
\varphi_{\Omega}(x)e^{-i\Omega t}$, while $\varphi$ solves the
stationary-state equation

\begin{equation}
-\frac{d^2\varphi_{\Omega}}{dx^2}+\left[U(x)-\Omega\right]\varphi_{\Omega}=0;
\label{sta_state_equat}
\end{equation}

$G_{\omega}^+$ is the "outgoing" Green function of equation
\ref{sta_state_equat} (but with a general $\omega$ instead of
$\Omega$).

The perturbation can be regarded as the force that is activated
upon the particle for a finite period $\tau$ ($f$ has momentum
dimensions, therefore $f/\tau$ has dimensions of force).

A straightforward substitution of eq.\ref{seek_solution} in
eq.\ref{Schro_eq} reveals the integral equation \cite{Azbel}

\begin{equation}
\tilde{f}(\omega-\Omega)\varphi_{\Omega}(0)+a_\omega-\int
d\omega'\tilde{f}(\omega-\omega')a_{\omega'}G_{\omega'}^{+}(0)=0
\label{int_eq}
\end{equation}
where
\begin{equation}
\tilde{f}(\omega)\equiv (2\pi)^{-1}\int dt f(t/\tau)\exp(i\omega
t) \label{fourier}
\end{equation}

and some tedious calculations \cite{Merzbacher} reveal that, for a
rectangular barrier with potential height $U$ and width $L$,

\begin{equation}
G_{\omega}(0)=-\frac{\coth \left[ \kappa L+i \arctan (k/\kappa)
\right]}{2\kappa} \simeq -(2\kappa)^{-1} \label{G0}
\end{equation}

where $k\equiv\sqrt{\omega}$ and $\kappa\equiv\sqrt{U-\omega}$ and
the approximation on the right hand side takes place for an opaque
barrier $\left(\sqrt{U}L\gg 1\right)$.

It should be noted that the results of this paper are valid for
any opaque barrier. We have chosen a rectangular barrier since it
has a relatively simple expression. In fact, the approximation in
eq.\ref{G0} (which is a good estimation for any opaque barrier)
yields, for all practical purposes, the same results.

In general, eq.\ref{int_eq} does not have an analytical solution.
Thus, we will begin with the numerical results. In Fig. 2 we
present the results of the exact solution of eq.\ref{int_eq} for
the temporally confined perturbation
\begin{equation}
f(t/\tau)=(f_0/\tau)\exp(-|t/\tau|) \label{temp-pert}.
\end{equation}

The $x$-axis represents the frequency $\omega$ (i.e., the
frequency of the transmitted particles), and the y-axis represents
the logarithm of $\tau$ (the perturbation's characteristic time).
Within this matrix, the gray levels indicate the logarithm of the
transmission coefficient, i.e.,$\ln|a_{\omega}G_{\omega}^+(x>L)|$.
In the figure, dark spots indicate a high probability of a
particle being emitted with the corresponding frequency, or energy
($\omega$).

From the figure we see that, for very large $\tau$ (i.e., the
adiabatic case), the outgoing frequency is almost identical to the
incoming one, $\Omega$. That is, the particles simply experience
spectral widening.

When $\tau$ decreases (and thus the process becomes more
energetic), the spectral width of the outgoing particles
increases, as it should by the uncertainty principle. The strange
part comes later, when $\tau$ reaches the value $\tau_0$ , after
which the increase in the spectral width becomes more moderate.
This is the point at which the activated particles prefer to be
activated to the first resonant energy $\simeq U+(\pi/2L)^2$.
These resonant energies \emph{are} barrier-dependent; therefore,
one should expect differences in these resonances from one barrier
to the next. However, these resonances have nothing to do with the
following effect, which is the main issue in this paper.

When $\tau$ decreases even further, the spectral width is split
into discrete probable frequencies-like a fan. That is, the
transmitting (through the barrier) particles are activated to
specific preferred frequencies, while \emph{activation to others
frequencies is forbidden}.

To explain this behavior, we can use the approximation that
$U-\Omega\gg\tau^{-1}$. Since, $U-\Omega$ can be arbitrarily
large, this approximation still enables us to discuss relatively
rapid changes. Within this temporal regime, only the spectral
vicinity of $\Omega$ affects the amplitude $a_\omega$. Thus, we
can approximate the Green function in eq. \ref{int_eq} according
to: $G_\omega^+(0) \rightarrow G_\Omega^+(0)$. The important
thing, which allows for this exchange, is that when $\Omega$ is
very far from $U$, $G_\omega^+(0)\simeq -(2\sqrt{U-\omega})^{-1}$
does not change much within the spectral range (again, recall that
this approximation is almost barrier-independent). Hence, with
this assumption in hand eq.\ref{int_eq} can readily be solved with
a Fourier transform:

\begin{equation}
A(t)=\frac{f(t/\tau)\varphi_\Omega(0)\exp(-i\Omega
t)}{1-f(t/\tau)G_\Omega^+(0)} \label{sol_for_A(t)}
\end{equation}

\begin{figure}
\psfig{figure=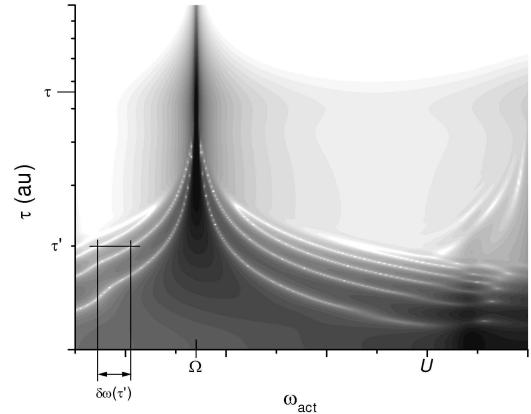,width=10cm,bbllx=80bp,bblly=45bp,bburx=700bp,bbury=478bp,clip=}
\caption{The probability of a particle being activated to energy
$\omega_{act}$, for various characteristic times $\tau$. The
y-axis represents $\tau$ on a logarithmic scale and the x-axis
represents the outgoing activation energy $\omega_{act}$. Inside
the figure, the darker the spot the higher the probability it
represents (again on a logarithmic scale). For an arbitrary
$\tau'$, the corresponding gap between adjacent forbidden
activation energies, i.e., $\delta\omega(\tau')$ is
illustrated.}\label{fig2}
\end{figure}

where $A(t)$ is the inverse Fourier counterpart of $a_\omega$. For
convenience, we use the dimensionless function $g$:
$f(t/\tau)\equiv(f_0/\tau)g(t/\tau)$ (where $f_0\equiv f(0)$). The
results in Fig.2, for example, were taken for $g(x)=\exp(-|x|)$.
It is assumed that $f_0 \equiv f(0)$ is the maximal value of the
perturbation. Therefore,

\begin{equation}
A(t)=\frac{f_0 g(t/\tau)\varphi_\Omega(0)\exp(-i\Omega
t)}{\tau-f_0 g(t/\tau)G_\Omega^+(0)} \label{sol_for_A(t)g}
\end{equation}

Eq.\ref{sol_for_A(t)g} determines two time scales

\begin{equation}
\tau_1 \equiv f_0G_\Omega \label{tau1}
\end{equation}
and
\begin{equation}
\tau_2 \equiv \tau g^{-1}(-f_0G_\Omega/\tau) \label{tau2}.
\end{equation}

When $\tau \gg \tau_1$, the denominator of eq. \ref{sol_for_A(t)g}
can be ignored, and the broadening of the spectral width is
exactly equal to the spectral width of the perturbation. Note that
since we are dealing with a strong perturbation (large $f_0$),
this characteristic time is very long.

When $\tau<\tau_1$ (and this is the more interesting regime), the
denominator cannot be ignored, and here $\tau_2$ comes into play:
when $t \ll \tau_2$, $A(t)\simeq \varphi_\Omega (0)\exp(-i\Omega
t)/G_\Omega^+(0)$; except for the oscillating term the absolute
value of $A(t)$ is almost a constant. However, when $t \gg
\tau_2$, the denominator of eq. \ref{sol_for_A(t)g} can again be
ignored (because of the rapid decay of $g$). Therefore, $A(t)$ has
a kind of rectangular shape: it is almost a constant for a time
$\sim \tau_2 \gg \tau$ and then it abruptly decays with the time
scale $\tau$ For such a "rectangular" shape, the Fourier transform
is something like the sinc function, which is oscillating and
decaying at the same time. This oscillating "frequency" of
$a_\omega$ is $\tau_2$ (note that we are dealing here with the
frequency space, hence the frequency has dimensions of time).
Thus, one can write $a_\omega \simeq a[(\omega-\Omega)\tau_2]$.
Multiplication by the Green function $G_\omega^+ (x>L)$ distorts
the picture a little (it increases the high frequencies'
contributions at the expense of those of the low frequencies), but
it does not change the fact that the particles can be activated
("act") only by a product of an integer and the quantity $\delta
\omega \sim \tau_2^{-1}$. That is,

\begin{equation}
\omega_{act}^n \simeq \Omega+n\delta\omega, \label{act_ener}
\end{equation}

where $n$ is an integer. Similarly, no activation occurs to the
forbidden ("for") activation levels

\begin{equation}
\omega_{for}^n \simeq
\Omega+\left(n+\frac{1}{2}\right)\delta\omega. \label{for_ener}
\end{equation}

These are the forbidden frequencies we were looking for. From the
definition of $\tau_2$ (eq. \ref{tau2}), one learns that it
depends not only on the characteristic time $\tau$ and on the
incoming frequency $\Omega$, but also on the strength of the
perturbation $f_0$ and, maybe more surprisingly, on the
\emph{functional behavior} of $f$. For example, suppose

\begin{equation}
g(x)=\exp(-|x|^m) \label{expg(x)}
\end{equation}

then

\begin{equation}
\tau_2=\tau\left[ \ln \left(-\frac{f_0G_\Omega}{\tau} \right)
\right]^{1/m} \label{dugma_tau2}
\end{equation}

It is clear that when the perturbation is strong ($f_0$ is large)
the dependence on $m$ can be very significant.

Let us take the following temporal dependence that corresponds to
a milder decay:

\begin{equation}
g_n(x)=\frac{1}{1+|x|^n} \label{g_n(x)}
\end{equation}

Such temporal behavior leads to a particularly simple solution,
which allows tracing the onset of the "discrete activation level
effect". In this case

\begin{equation}
a_\omega=\frac{1}{2\pi}\int dt A(t)\exp(i\omega
t)=\frac{f_0}{2\pi}\int dx \frac{\exp(ixs)}{|x|^n+\alpha^n}
\label{a_omega}
\end{equation}

where we adopt the dimensionless parameters $s\equiv
(\omega-\Omega)\tau$, $\alpha^n \equiv -G(0)f_0/\tau+1$, and of
course $x \equiv t/\tau$.

But eq.\ref{a_omega} can also be rewritten

\begin{equation}
a_\omega=\frac{f_0}{\alpha^{n-1}} \tilde{g}_n (s\alpha)
\label{a_omega2}
\end{equation}

where $\tilde{g}_n$ is the inverse Fourier transform of $g_n$ (see
eq.\ref{fourier}).

For $y \rightarrow \infty$, an asymptotic expression can be
written

\begin{equation}
\tilde{g}_n(y) \sim \cos[y\cos(\pi/n)]\exp[-y\sin(\pi/n)],
\label{asimp_for_g}
\end{equation}

(for $n=2$ and $n=4$ this is the only term in the asymptotic
expansion).

Thus, the oscillating part of $a_\omega$ reads

\begin{equation}
\cos \left[ (\omega-\Omega) \tau \left(-\frac{Gf_0}{\tau}+1
\right)^{1/n} \cos \left(\frac{\pi}{n} \right) \right]
\label{osci_part_of_a_w}
\end{equation}

Similar to the exponential case (eq. \ref{expg(x)}) we can define
a characteristic time

\begin{equation}
\tau_2 \sim \tau (-Gf_0/\tau+1)^{1/n}\cos(\pi/n)
\label{char_time_tau2}
\end{equation}

which determines the activation energies according to eq.
\ref{act_ener}. Similar to eq. \ref{dugma_tau2}, $\tau_2
\rightarrow \tau$ when the \emph{functional} behavior of the
perturbation dies out very quickly ($n \rightarrow \infty$).
Moreover, eq. \ref{char_time_tau2} also reveals the onset of the
discrete activation pattern. The parameter $n$ \emph{must be
larger than 2}, otherwise no oscillations occur.

Therefore, we conclude that in order to obtain the discrete
activation level effect, the perturbation must die out faster than
$(t/\tau)^{-2}$.

It should be noted, that because of the point-like nature of the
alternating impurity, the whole derivation of the effect depends
on the barrier via the Green function, however, as approximation
(\ref{G0}) suggests, any opaque barrier would yield similar
results.

To summarise, the activation energy in the case of a strongly
varying potential barrier was investigated. It was shown that when
the temporal perturbation amplitude is very large, the particles
cannot be activated to certain discrete energies
$\omega_{for}^m\simeq\Omega+(m+1/2)\delta\omega$, where
$\delta\omega \sim [\tau g^{-1}(-f_0G_\Omega/\tau)]^{-1}$, $\tau$
is the characteristic time of the perturbation, $f_0$ is its
strength and $g^{-1}$ is the inverse function of the
perturbation's temporal dependence. Moreover, it was demonstrated
that in order to see this forbidden activation effect, the
perturbation must die out faster than $t^{-2}$ .

I would like to thank Prof. Mark Azbel for stimulating
discussions.


\begin{references}

\bibitem{Zangwill} A. Zangwill and P. Soven, Phys. Rev. Lett., 45, 204 (1980)

\bibitem{Ivlev_Melnikov} B.I. Ivlev and V.I. Mel'nikov, Phys.Rev. Lett. 55,1614 (1985)

\bibitem{Fisher} M.P.A. Fisher, Phys. Rev. B, 37, 75 (1988)

\bibitem{Azbel} M. Azbel, Phys. Rev. B, 43, 6847 (1991);
Phys. Rev. Lett. 68, 98 (1992); Europhys. Lett. 18, 537 (1992)

\bibitem{Aleiner_Andreev} I.L. Aleiner and A.V. Andreev, Phys. Rev. Lett. 81, 1286 (1998)

\bibitem{Merzbacher} E. Merzbacher, Quantum Mechanics, (Wiley 1970)

\end{references}
\end{document}